\documentclass[a4paper,11pt]{article}

\usepackage{jheppub}
\usepackage{epsfig,graphicx}
\usepackage{longtable}
\usepackage{amssymb}
\usepackage{amsmath}
\usepackage{hyperref}
\usepackage{orgm}

\title{The Evolution of Meson Masses in a Strong Magnetic Field}

\author[a]{M.A.Andreichikov}
\author[a,b,c]{B.O.Kerbikov }
\author[a,b]{E.V.Luschevskaya }
\author[a]{Yu.A.Simonov }
\author[a]{O.E.Solovjeva }

\affiliation[a]{Institute for Theoretical and Experimental Physics, B. Cheremushkinskya 25, 117218 Moscow, Russia}
\affiliation[b]{Moscow Institute of Physics and Technology, Institutskiy per. 9, 141700 Dolgoprudniy, Russia}
\affiliation[c]{Lebedev Physical Institute RAS, Leninsky prospekt 53, Moscow, Russia}

\emailAdd{andreichicov@mail.ru}
\emailAdd{borisk@itep.ru}
\emailAdd{luschevskaya@itep.ru}
\emailAdd{simonov@itep.ru}
\emailAdd{olga.solovjeva@itep.ru}

\abstract{
Spectra of $q \bar{q}$ hadrons are investigated in the framework of the
 Hamiltonian  obtained from the relativistic path integral in external homogeneous magnetic field.
The spectra of all 12 spin-isospin s-wave states, generated by  $\pi$ and
$\rho$ mesons with different spin projections, are studied both analytically
and numerically on the  lattice as functions of (magnetic field) $eB$. Results
are in agreement  and demonstrate three types of behavior, with characteristic
splittings  predicted by the theory.}

\begin{document}

\newcommand{\beqn}{\begin{eqnarray}}
\newcommand{\eeqn}{\end{eqnarray}}
\newcommand{\eq}[1]{(\ref{#1})}
\newcommand{\lr}[1]{ \left( #1 \right) }
\newcommand{\lrs}[1]{ \left[ #1 \right] }
\newcommand{\lrc}[1]{ \left\{ #1 \right\} }
\newcommand{\vev}[1]{ \langle \, #1 \, \rangle }
\newcommand{\Tr}{ {\rm Tr} \, }
\newcommand{\tr}{ {\rm Tr} \, }
\newcommand{\re}{ {\rm Re} \, }
\newcommand{\im}{ {\rm Im} \, }
\renewcommand{\Re}{ {\rm Re} \, }
\renewcommand{\Im}{ {\rm Im} \, }
\newcommand{\rvac}{ \, | 0 \rangle }
\newcommand{\lvac}{ \langle 0 | \, }
\newcommand{\sign}{ {\rm sign} \,  }
\newcommand{\logo}{\\ \vskip -15mm \leftline{\includegraphics[scale=0.3,clip=false]{logo.eps}} \vskip 7mm}

\newcommand{\const}{\mbox{const}}
\newcommand{\fm}{\mbox{fm}}
\newcommand{\Mev}{\mbox{MeV}}
\newcommand{\Gev}{\mbox{GeV}}

\newcommand{\Tl}{\mbox{Tl}}
\newcommand{\phys}{\mbox{phys}}
\newcommand{\latt}{\mbox{lat}}
\newcommand{\ren}{\mbox{ren}}
\newcommand{\plq}{\mbox{pl}}
\newcommand{\rt}{\mbox{rt}}

\newcommand{\be}{\begin{equation}}
\newcommand{\ee}{\end{equation}}
\def\la{\mathrel{\mathpalette\fun <}}
\def\ga{\mathrel{\mathpalette\fun >}}
\def\fun#1#2{\lower3.6pt\vbox{\baselineskip0pt\lineskip.9pt
\ialign{$\mathsurround=0pt#1\hfil ##\hfil$\crcr#2\crcr\sim\crcr}}}
\newcommand{\veX}{\mbox{\boldmath${\rm X}$}}
\newcommand{{\SD}}{\rm SD}
\newcommand{\pp}{\prime\prime}
\newcommand{\veY}{\mbox{\boldmath${\rm Y}$}}
\newcommand{\vex}{\mbox{\boldmath${\rm x}$}}
\newcommand{\vey}{\mbox{\boldmath${\rm y}$}}
\newcommand{\ver}{\mbox{\boldmath${\rm r}$}}
\newcommand{\vesig}{\mbox{\boldmath${\rm \sigma}$}}
\newcommand{\vedelta}{\mbox{\boldmath${\rm \delta}$}}
\newcommand{\veP}{\mbox{\boldmath${\rm P}$}}
\newcommand{\vep}{\mbox{\boldmath${\rm p}$}}
\newcommand{\veq}{\mbox{\boldmath${\rm q}$}}
\newcommand{\veK}{\mbox{\boldmath${\rm K}$}}
\newcommand{\vez}{\mbox{\boldmath${\rm z}$}}
\newcommand{\veS}{\mbox{\boldmath${\rm S}$}}
\newcommand{\veL}{\mbox{\boldmath${\rm L}$}}
\newcommand{\vem}{\mbox{\boldmath${\rm m}$}}
\newcommand{\veQ}{\mbox{\boldmath${\rm Q}$}}
\newcommand{\vel}{\mbox{\boldmath${\rm l}$}}
\newcommand{\veR}{\mbox{\boldmath${\rm R}$}}
\newcommand{\veA}{\mbox{\boldmath${\rm A}$}}
\newcommand{\ves}{\mbox{\boldmath${\rm s}$}}
\newcommand{\vek}{\mbox{\boldmath${\rm k}$}}
\newcommand{\ven}{\mbox{\boldmath${\rm n}$}}
\newcommand{\veu}{\mbox{\boldmath${\rm u}$}}
\newcommand{\veh}{\mbox{\boldmath${\rm h}$}}
\newcommand{\vew}{\mbox{\boldmath${\rm w}$}}
\newcommand{\verho}{\mbox{\boldmath${\rm \rho}$}}
\newcommand{\vexi}{\mbox{\boldmath${\rm \xi}$}}
\newcommand{\veta}{\mbox{\boldmath${\rm \eta}$}}
\newcommand{\veB}{\mbox{\boldmath${\rm B}$}}
\newcommand{\veH}{\mbox{\boldmath${\rm H}$}}
\newcommand{\veE}{\mbox{\boldmath${\rm E}$}}
\newcommand{\veJ}{\mbox{\boldmath${\rm J}$}}
\newcommand{\veal}{\mbox{\boldmath${\rm \alpha}$}}
\newcommand{\vegam}{\mbox{\boldmath${\rm \gamma}$}}
\newcommand{\vepar}{\mbox{\boldmath${\rm \partial}$}}
\newcommand{\vepi}{\mbox{\boldmath${\rm \pi}$}}
\newcommand{\vemu}{\mbox{\boldmath${\rm \mu}$}}
\newcommand{{\Mc}}{\mathcal{M}}
\newcommand{\llan}{\langle\langle}
\newcommand{\rran}{\rangle\rangle}
\newcommand{\lan}{\langle}
\newcommand{\ran}{\rangle}


\maketitle
\flushbottom

\section{Introduction}
The influence of magnetic field (MF) on the strong interacting particles is an
actively  discussed topic, see, e.g. a recent review \cite{1}. When MF is not
ultra-intense ($eB \ll \sigma$, where $\sigma = 0.18 \ GeV^2$ is a confinement
string tension)\footnote{We use relativistic system of units $\hbar = c= 1$, then $1 \ GeV^2 = 5.12 \cdot 10^{19} \ G$}, the main characteristics related to the behavior of hadrons in
MF are magnetic moments and magnetic susceptibilities, while for the strong MF
limit ($eB \geq \sigma$) the hadron  energy  and width depend on  MF
directly.

These topics are important in astrophysics of neutron stars \cite{2}, in
cosmological theories \cite{3}, in atomic physics \cite{30,31,31a,34} in the physics of heavy ion collisions
\cite{4}, and in the high-intensity lasers \cite{5}.

On the theoretical side the main directions of research in this area are the
lattice studies \cite{6}-\cite{13}, \cite{39}-\cite{45}, the chiral Lagrangians with MF
\cite{14}-\cite{17}, effective hadron Lagrangians \cite{1,19},\cite{46}-\cite{49}, and recently
developed path integral Hamiltonians (PIH) \cite{20}-\cite{23}, and the chiral
Lagrangian with quark degrees of freedom \cite{24}.

The PIH method has appeared to be well suited to the inclusion of an arbitrary
external MF.
 Here one obtains simple expressions for magnetic
  moments of hadrons, mesons \cite{25} and baryons \cite{26}, which are in a
   good agreement with available experimental and lattice data,  as well as
    with existing model calculations. We stress at this point, that in all calculations
      done within the PIH framework, the final results are expressed in
    terms of basic QCD parameters - string tension $\sigma$, $\alpha_s$ and current quark masses $m_q$.

A sample of light neutral meson masses in MF (actually, the meson energies for
zero longitudial momentum) has been calculated with PIH framework in
\cite{20,21,211}, and the Nambu-Goldstone (NG) modes in MF have been studied in
\cite{24}. In all cases the resulting values of $M_i(B)$ are in reasonable
agreement with lattice data from \cite{8}-\cite{10}. The three-body neutral
systems in strong MF were studied with PIH in \cite{27}, but there is no
lattice data   now to compare with.

In a general case, solving the spectral problem  for hadrons in MF is a
cumbersome task. To proceed   with analytic calculations, one should use some
special techniques. One of them is the Pseudomomentum approach. It was
introduced in \cite{28} to separate center-of-mass (c.m.) motion from the
relative motion in the nonrelativistic Hamiltonian for the neutral system in
MF. This approach was extended to  the relativistic sector  in the   PIH
framework for two-body systems in \cite{20,211,21} and for three-body systems
in \cite{29}. The Pseudomomentum approach  is applicable  only for electrically
neutral systems, and for the charged ones  an exact analytical answer was
obtained  only in an unphysical model of charged meson with equally charged
quark constituents  \cite{211}.

 Below we are suggesting a new approximate analytic
method of constituent separation (CS) that allows to get  a quantitative result
for any meson masses  with 15\% accuracy for the strong MF ($eB \gg \sigma$)
and with 20\% accuracy for $eB < \sigma$.  As will be shown, the CS method
allows to study charged and neutral systems in the same way. To introduce it,
we first write the relativistic Hamiltonian   in MF within PIH formalism and
exploit the  oscillator representation for the confinement interaction
used before in \cite{20,211,21} with 5\% accuracy. This   allows to split
the Hamiltonian into transversal and longitudial (with respect  to  the MF
direction) parts analytically. All the rest interaction - one-gluon exchange,
spin-dependent and self-energy interactions are studied perturbatively.

 Our final results for the neutral mesons in MF are obtained in two
 independent ways: via Pseudomomentum  and the CS methods, which allows to
 check the accuracy of our results.

 The paper is organized as
follows: in Section 2 we write relativistic Hamiltonian and discuss the main
features of CS method. (Details of this method are  discussed in Appendix A).
As  a result   we obtain in Section 2    the hadron mass and the ground  state
wave function  as a function of $eB$ and $\sigma$  for an arbitrary meson in
MF. In Section 3   a classification  of meson mass trajectories  with different
spin and isospin projections is given with the corresponding asymptotics in
high MF regime $eB \rightarrow \infty$. In Section 4   the perturbative
correction  due to the one-gluon exchange is calculated  and the  absence  of
the color Coulomb collapse   is   demonstrated. The CS wave function for neutral
mesons is discussed in in the Appendix B. In Section 5   the spin-spin
interaction in MF and the  seemingly possible ``hyperfine collapse'' is discussed.
 In Section 6  a general discussion of the we
  spin-isospin splitting is given.
In Section 7 we study  the chiral and nonchiral treatment of pion masses in MF.
  In Section 8   the details of our lattice calculations are given.
  Results of both analytic and lattice results are discussed in the  concluding
   Section 9.

\section{The relativistic Hamiltonian of quark systems}
We start from the relativistic  Hamiltonian of the N-quark system in an
external homogeneous MF, which according to \cite{20}-\cite{22} is
\begin{equation}
  \label{eq:1}
   H_0= \sum^N_{i=1} \frac{(p^{(i)}_k - e_i A_k)^2 + (m^q_i)^2 + \omega^2_i - e_i \vesig_i \veB}{2\omega_i},
\end{equation}
where $\omega_i$ are  virtual quark energies to be intergrated over in the path
integral, and $m^q_i$ are current quark masses. At this step we neglect any
internal interactions between quarks, i.e. confinement, gluon-exchange, etc. It
is convenient to choose symmetrical gauge for MF $\veA_i = \frac12 (\veB \times
\ver_i)$ which  allows to define an angular momentum projection $m_i$ for each
quark as a quantum number. The  spectrum  of  (\ref{eq:1}) with $m_i = 0$ is
\begin{equation}
  \label{eq:2}
  \varepsilon_i (\omega_i) = \frac{(m^q_i)^2 +\omega^2_i + |e_i|B (2n_{\bot i} + 1) - e_i \vesig_i \veB+ (p_z^{(i)})^2}{2\omega_i}.
\end{equation}
According to \cite{20}-\cite{22}  the  physical spectrum is given by the
stationary point value of $\varepsilon$, with respect to $\omega_i$

\begin{equation}
  \label{eq:3}
  \begin{split}
 & \left.\frac{d\varepsilon_i (\omega_i)}{d\omega_i}\right|_{\omega_i=\omega_i^{(0)}} = 0,\ \varepsilon^{(i)} (\omega_i^{(0)}) \equiv \bar \varepsilon^{(i)},\bar E_0 \equiv \sum_{i=1}\bar\varepsilon_i,\\
& \bar \varepsilon_i = \sqrt{(m^q_i)^2 + (p^{(i)}_z)^2 +  |e_i| B(2n_\bot +1)- e_i \vesig_i \veB} .
  \end{split}
\end{equation}
It is easy to see that this spectrum coincides with the solution of the Dirac
equation for  N non-interacting relativistic particles  in  MF.

 As in \cite{211} we  now introduce the
confining interaction $V_{\rm conf}$,  which is treated nonperturbatively,
while the other interactions like one-gluon exchange $V_{OGE}$, spin-dependent
interaction $a_{SS}$ and self-energy corrections $\Delta M_{SE}$ are treated perturbatively in the next
sections. The Hamiltonian becomes
\begin{equation}
  \label{eq:4}
  H_d = H_0 + V_{\rm conf}
\end{equation}
with the ground state eigenvalue   $M_d$ (nonperturbative, or dynamical mass)  and the ground state wave function  $|\Psi_0
\rangle$. The total meson mass is a sum of $M_d$ and the perturbative corrections
\begin{equation}
  \label{eq:5}
  M_{\rm total} = M_d + \langle \Psi_0|V_{OGE}|\Psi_0 \rangle + \langle a_{SS} \rangle + \Delta M_{SE}.
\end{equation}
One can note, that the contribution of the $V_{\rm conf}$ in strong MF ($eB \gg \sigma$) is
negligible in the plane transverse to the MF direction and should be retained
only for lowest levels, which we call ``zero hadron states" (ZHS) (see below).
Another feature is that in strong MF regime the translational invariance of the
center-of-mass (c.m.) is broken due to magnetic forces (each quarks is placed
on its own Landau level), but the confinement   still defines the motion  of
quarks in the direction along the MF.

To simplify calculations  we chose the confining term in the variable quadratic
form \cite{211,21}, restoring its original linear form at the stationary point (it was
checked   to be accurate within about 5\%), namely
\begin{equation}
  \label{eq:6}
  V_{\rm conf}^{(q\bar q)} = \sigma|\ver_1-\ver_2|\to \frac{\sigma}{2\gamma} (\ver_1-\ver_2)^2 + \frac{\sigma\gamma}{2},
\end{equation}
where $\gamma$ is variational parameter and $\sigma = 0.18 \ GeV^2$ is a confinement string tension. The dependence of the string tension $\sigma$ on the MF is caused by the fluctuating $q\bar{q}$ pairs embedded to the string and provides a correction about $\frac{\Delta \sigma}{\sigma} \sim 15 \%$ at $eB \sim 1 \ GeV^2$. This phenomenon was studied on the lattice in \cite{21-1} and was confirmed within PIH formalism in \cite{21-2}. The correction to the ground state caused by this effect is beyond the declared accuracy and is neglected in what follows. To produce an approximation for the
energy, one should minimize the resulting  state energy obtained from the Hamiltonian (\ref{eq:4}) with
respect to $\omega_i$ and $\gamma$ simultaneously.

The oscillator approximation (\ref{eq:6}) gives an advantage to separate motion along the $z$ axis(parallel to the MF) and in $x-y$ plane
\begin{equation}
  \label{eq:7}
  \Psi_0 = \psi^{(z)}(z^{(1)},z^{(2)})\psi^{(\perp)}(\ver_{\perp}^{(1)},\ver_{\perp}^{(2)});\ H_d = H_{\perp} + H_3,
\end{equation}
where the motion along the $z$-axis is defined by the Hamiltonian
\begin{equation}
  \label{eq:8}
 H_3 = \left(\frac{(p^{(1)}_3)^2}{2\omega_1} + \frac{(p^{(2)}_3)^2}{2\omega_2} +
  \frac{\sigma}{2\gamma}(z^{(1)} - z^{(2)})^2\right) \rightarrow 
   \frac{P^2_3}{2(\omega_1 + \omega_2)} + \frac{\pi^2_3}{2\tilde \omega} + \frac{\sigma}{2\gamma}
   \eta^2_3,
\end{equation}
where we use c.m. reference frame with $P_3 = p_3^{(1)} + p_3^{(2)}, ~~
 \eta_3 = z^{(1)} -z^{(2)}; ~~ \pi_3 = \frac{1}{i} \frac{\partial}{\partial \eta_3}, ~~ \tilde \omega =
\frac{\omega_1\omega_2}{\omega_1+\omega_2}$.The longitudial part of the ground state energy is 
\begin{equation}
  \label{eq:9}
  M_3^0 =\frac{P^3_3}{2(\omega_1 +\omega_2)} + \left( n_3 + \frac12\right) \sqrt{\frac{\sigma}{\tilde \omega\gamma}};\ n_3 = 0;\ P_3 = 0.
\end{equation}
For the motion in the transversal plane one can use an approximation of decoupled quarks at large MF, making the following substitution
\begin{equation}
  \label{eq:10}
  (\ver_{\perp}^{(1)} - \ver_{\perp}^{(2)})^2 = (\ver_{\perp}^{(1)} -\ver_{\perp}^0)^2
  + (\ver_{\perp}^{(2)} - \ver_{\perp}^0)^2 - 2 (\ver_{\perp}^{(1)} - \ver_{\perp}^0) (\ver_{\perp}^{(2)} - \ver_{\perp}^0) \rightarrow \sum_{i=1}^2(\ver_{\perp}^{(i)} - \ver_{\perp}^0)^2,
\end{equation}
where c.m. position $\ver_{\perp}^0$ is fixed at the origin in $x-y$ plane. This approximation corresponds to the configuration where the confinig string connects each quark to the c.m., i.e. the string is effectively elongated. The magnetic energy of each quark in strong MF (Landau level) is larger than the confinig interaction with the factor $\frac{eB}{\sigma}$, which make this approximation legitimate at $eB > \sigma$ regime. To extend our method to the $eB < \sigma$ region, where the behaviour is mostly defined by confinement, one should introduce an effective sting tension $\sigma_1$ and $\sigma_2$ for each part of the string, connecting quarks to the c.m.
\begin{equation}
  \label{eq:11}
  V_{\rm conf} = \frac{\sigma_1}{2\gamma}
  (\ver_{\perp}^{(1)} - \ver_{\perp}^0)^2 + \frac{\sigma_2}{2\gamma}(\ver_{\perp}^{(2)} - \ver_{\perp}^0)^2 + \frac{\sigma \gamma}{2},
\end{equation}
to compensate an effective string elongation. As shown in Appendix A, the appropriate values of $\sigma_1, \sigma_2$ are 
\begin{equation}
  \label{eq:12}
  \sigma_1 = \frac{\sigma}{1 + \frac{\omega_1}{\omega_2}}; \ \sigma_2 = \frac{\sigma}{1 +
  \frac{\omega_2}{\omega_1}}.
\end{equation}
Using this "$\sigma$-renormalization" procedure, one can show that the dynamical mass of the ground state $M_d$, calculated in \cite{211} with the Pseudomomentum technique for neutral mesons, exactly coincides with the dynamical mass obtained in the above CS formalism for the arbitrary value of MF.As a result this approximation make quarks effectively decoupled in $x-y$ plane and one can write 
\begin{equation}
  \label{eq:13}
  \psi^{(\perp)}(\ver_{\perp}^{(1)},\ver_{\perp}^{(2)}) = \psi^{(\perp)}_1(\ver_{\perp}^{(1)},)\psi^{(\perp)}_2(\ver_{\perp}^{(2)}). 
\end{equation}
The transversal part of the hamiltonian $H_{\perp}$ has the ground state energy 
 \begin{equation}
   \label{eq:14}
    M_\bot^0 =\sum^2_{i=1} \frac{m^2_i +\omega^2_i - e_i \vesig_i \veB+ ( 2n_\bot^{(i)} +1) \sqrt{(e_i B)^2 + 4\sigma_i \omega_i/\gamma}}{2\omega_i};\ n_{\perp}^{(i)} = 0,
 \end{equation}
where $\sigma_i$ are given by (\ref{eq:12}). The total dynamical mass is given by the sum 
\begin{equation}
  \label{eq:15}
 M_d = M_\bot^0 + M_3^0 + \frac{\sigma\gamma}{2}.
\end{equation}
The actual trajectories for the dynamical mass in MF, $M_d (eB)$ are obtained
using the stationary point conditions in a similar way as (\ref{eq:3})
\begin{equation}
  \label{eq:16}
\tilde M_d = M_d^0 (\omega_i^{(0)}, \gamma^{(0)}),\ \left.\left.\frac{\partial M_d}{\partial\gamma}\right|_{~\gamma=\gamma^{(0)}}= \frac{\partial  M_d}{\partial \omega_i} \right|_{~\omega_i=\omega_i^{(0)}}=0.
\end{equation}
The corresponding wave function for the ground state $\tilde M_d$ is

\begin{equation}
  \label{eq:17}
  \Psi_0 = \left( \frac{\tilde{\omega}^{(0)} \Omega_z}{\pi} \right)^{\frac{1}{4}} \left( \frac{\omega_1^{(0)} \Omega_1 \omega_2^{(0)} \Omega_2}{\pi^2} \right)^{\frac{1}{2}} e^{-\frac{\omega_1^{(0)} \Omega_1}{2}(\vect{r}_1^{\perp})^2 -\frac{\omega_2^{(0)} \Omega_2}{2}(\vect{r}_2^{\perp})^2 -\frac{\tilde{\omega}^{(0)} \Omega_z}{2}(\eta_z^2)},
\end{equation}

where $\Omega_1$,  $\Omega_2$ and  $\Omega_z$ are harmonic oscillator frequences
\begin{equation}
  \label{eq:17-2}
  \Omega_i = \frac{1}{2\omega_i^{(0)}} \sqrt{(e_iB)^2 + \frac{4\sigma_i \omega_i^{(0)}}{\gamma^{(0)}}};\ \Omega_z = \sqrt{\frac{\sigma}{\tilde{\omega}^{(0)} \gamma^{(0)}}}.
\end{equation}
Comparing (\ref{eq:17}) with the  same wave function obtained in \cite{211} for
neutral mesons one can see that now we have two elongated  ellipsoids for each
quark instead of one ellipsoid in $\gvect{\eta} = \ver_1 - \ver_2$,  but the resulting spectra coincide.

\section{Meson trajectories in MF}

We turn now to  the general structure of the meson  spectrum and the limits of
weak ($eB < \sigma$) and strong ($eB \gg \sigma$) MF.

 For small MF  both
$\gamma$ and $\omega$ are independent of MF in the leading order and the lowest
order the correction to the dynamical mass is
\begin{equation}
  \label{eq:18}
  \tilde{M}_d(B) = \tilde{M}_d(B=0) - \sum_{i=1}^2 \frac{e_i \vesig_i \veB}{2\omega_i^{(0)}} = \tilde{M}_d(B=0) - \vemu\veB + c|e\vect{B}|,
\end{equation}
 where $\vemu$  is the magnetic moment  of the hadron, and the $c|e \vect{B}|$ term is c.m. energy contribution (the lowest Landau level) in MF for the charged mesons (note,
  that in this paper we discuss only s-wave hadrons and all orbital momenta are zero).

Magnetic moments in PIH formalism have been calculated in \cite{25} for mesons
and are in good agreement with experiment and  available lattice data. It is
easy to see there that for massless quarks the expansion in (\ref{eq:18}) is actually done in
powers of $ \left( \frac{eB}{\sigma} \right)$.

 For the strong MF limit the
situation is more complicated.  Confining ourselves to the lowest Landau levels
(LLL) for all quarks and antiquarks, i.e. $n_{\perp}^{(i)} = 0$ in
(\ref{eq:9}) and (\ref{eq:14}), we can separate out the hadrons, which consist
of only LLL states of both quarks with $e_i \sigma_z^i = |e_i|,\ i = 1,2$.
These states  are MF-independent at $eB \rightarrow \infty$ and we shall call
them ``zero hadron states" (ZHS). Note, that ZHS do not possess definite total
spin and isospin quantum numbers.

 All other hadron states, except for ZHS, will have
energies growing with MF as $\sim \sqrt{|eB|}$ and therefore thermodynamically
suppressed at large MF. In the limit of strong MF the dynamical masses for ZHS
can be written as
\begin{equation}
  \label{eq:19}
  M_d^{(ZHS)}(eB \gg \sigma) \simeq M_3^0 + \sum_{i=1}^2 \frac{m_i^2 +
  \omega_i^2}{2\omega_i} + \frac{\sigma \gamma}{2}.
\end{equation}
The stationary point analysis according to (\ref{eq:16}) for $m_1 = m_2 = 0,\
e_i \sigma_z^i = 1,\ i = 1,2$, yields
\begin{equation}
  \label{eq:20}
  \begin{split}
  \omega^{(0)} = & \omega^{(0)}_1 = \omega^{(0)}_2 = \frac{\sqrt{\sigma}}{2};\ \gamma^{(0)} = \frac{1}{\sqrt{\sigma}};\\
 & \tilde  M_d^{ZHS} = 2\sqrt{\sigma} = 4\omega^{(0)}.
  \end{split}
\end{equation}
The same result was obtained in \cite{20,211, 21,27} with the Pseudomomentum technique.

We turn now to the meson states still with zero orbital momentum and not
belonging to the ZHS states, i.e. violating the equality
$\frac{e_i}{|e_i|}\sigma_z^i = 1$. The resulting meson energy according to
(\ref{eq:15})-(\ref{eq:16}) for $m_1 = m_2 = 0,\ P_3 = 0,\ e_1 = |e_1|, e_2 =
|e_2|$ is
\begin{equation}
  \label{eq:21}
  M_d = M_{\perp}^0 + M_3^0 + \frac{\sigma \gamma}{2} = \frac{\omega_1}{\gamma} + \frac{\sigma}{\gamma e_1 B} + \frac{\sigma}{\gamma e_2 B} + \frac{1}{2} \sqrt{\frac{\sigma}{\tilde \omega \gamma}} + \frac{\omega_2^2 + 2e_2B}{2\omega_2} + \frac{\sigma \gamma}{2},
\end{equation}
which yields
\begin{equation}
  \label{eq:22}
  \begin{split}
  \omega_2^{(0)} = & \sqrt{2e_2B},\ \gamma^{(0)} = \frac{1}{\sqrt{2\sigma}},\ \omega_1^{(0)} = 2^{-5/6}\sqrt{\sigma},\\
 & \tilde M_d^{I} = \sqrt{2e_2B} + \sqrt{2\sigma}.
  \end{split}
\end{equation}
The same result  occurs when $\frac{e_1}{|e_1|}\sigma_z^1 = -1,\
\frac{e_2}{|e_2|}\sigma_z^2 = 1$ with replacement $ e_2 \rightarrow e_1$ Now we
turn to the case when both products $\frac{e_1}{|e_1|}\sigma_z^1$ and
$\frac{e_2}{|e_2|}\sigma_z^2$ are negative. In this case one obtains
\begin{equation}
  \label{eq:23}
  \begin{split}
&  \omega_i^{(0)} = \sqrt{2e_iB},\ \gamma^{(0)} = 2^{-2/3}(\sigma \tilde \omega^{(0)})^{-\frac{1}{3}},\\
&  \tilde M_d^{II} = \sqrt{2e_1B} + \sqrt{2e_2B} + \frac{3 \sigma^{2/3}}{2^{5/3}(\tilde \omega^{(0)})^{1/3}}.
  \end{split}
\end{equation}
Thus we have three different asymptotic modes for s-wave meson dynamical masses
$M_d$  in MF, classified with respect to spin projections
\begin{equation}
\label{eq:24}
\begin{split}
1)& \ ZHS: \ e_1 \sigma_z^1 >0,\ e_2 \sigma_z^2 >0:\\
 & \ \tilde M_d^{ZHS}(eB \gg \sigma) = 2 \sqrt{\sigma};\\
2)& \ I:   \ e_1 \sigma_z^1 >0,\ e_2 \sigma_z^2 <0:\\
 & \ \tilde M_d^{I}(eB \gg \sigma) = \sqrt{2e_1B} + \sqrt{2\sigma};\\
3)& \ II:  \ e_1 \sigma_z^1 <0,\ e_2 \sigma_z^2 <0:\\
 & \ \tilde M_d^{II}(eB \gg \sigma) = \sqrt{2e_1B} + \sqrt{2e_2B}.\\
\end{split}
\end{equation}

We shall return to this classification later  in Section 6  in our  study of
spin-isospin splittings in weak MF regime.

\section{One-gluon exchange in MF}

The first order perturbation correction for one-gluon exchange potential(OGE, or color Coulomb interaction)
in MF  entering in (\ref{eq:5}) according to  \cite{21}  is

 \begin{equation}
  \label{eq:25}
  V_{OGE} = - \frac{16 \pi \alpha_s^{(0)}}{3 \left[ Q^2
  \left( 1+ \frac{\alpha_s^0}{4\pi} \frac{11}{3}
   N_c \ln{\left(\frac{\vect{q}^2 + M_B^2}{\Lambda^2} \right)} \right)
   + \frac{\alpha_s^{(0)}n_f|eB|}{\pi} e^{-\frac{q_{\perp}^2}{2|eB|}}T \left(\frac{q_3^2}{4\sigma} \right) \right]},
\end{equation}

\begin{figure}[t]
  \begin{center}

    \includegraphics[width=.7\textwidth]{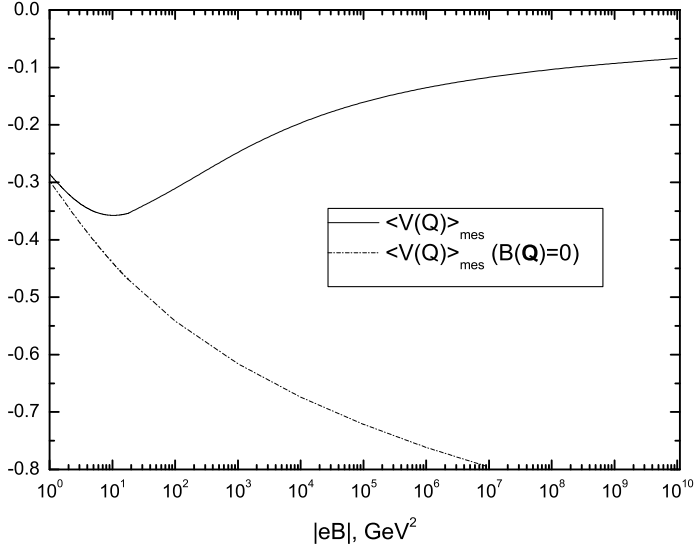}
    \caption{Color Coulomb matrix element $\langle V_{OGE} \rangle$ (in GeV) in MF with
    screening (solid line) and without screening (dashed line) by $q \bar q$ pairs. $\langle V_{OGE} \rangle$ saturates at $eB \rightarrow \infty$ and the fall-to-the-center phenomenon doesn't occur.  }
  \end{center}
\label{figAF}
  \end{figure}
where $N_c = 3,\ n_f = 2,\  \alpha_s^{(0)} = 0.42$, QCD parameter $\Lambda = 0.3\ GeV$, and the $M_B^2 = 2\pi\sigma = 1.1 \ GeV^2$ preventing Landau singularity was calculated in \cite{456}. Form (\ref{eq:25}) includes screening of the OGE  potential by
the  quark-antiquark pairs created in MF. This effect prevents the
``fall-to-the-center" phenomenon for ZHS hadrons in MF, as  shown in the
Fig.1 and 7. One can see that the matrix element $\langle \Psi | V_{OGE} |
\Psi \rangle$ for meson saturates  at $eB \sim 10 \ GeV^2$ and   the system
becomes ``asymptotically free" in $eB \rightarrow \infty$ limit when $\langle
\Psi | V_{OGE} | \Psi \rangle \rightarrow 0$. The driving force of the Coulomb
collapse is an uncontrollable growth of the Coulomb interaction when the system
is squeezed by MF forces. The role of screening  of the  Coulomb interaction in
MF has a long story, see e.g. \cite{30,31,31a} for atomic systems.

The next step is to average the potential (\ref{eq:25}) over the wave function
(\ref{eq:17}) obtained by the   CS method.
\begin{equation}
  \label{eq:26}
  \langle V_{OGE} \rangle = \langle \Psi_0 | V_{OGE} | \Psi_0 \rangle =  \int d^3 \vect{r}_1 d^3 \vect{r}_2 |\Psi_0(\vect{r}_1,\vect{r}_2)|^2 V_{OGE}(\vect{r}_1 - \vect{r}_2).
\end{equation}
Separating the integration in $x-y$ plane and  in $z$-direction, one has
\begin{equation}
  \label{eq:27}
  \langle V_{OGE} \rangle = \int d^2 \vect{r}_1^{\perp} d^2 \vect{r}_2^{\perp}
   d \eta_z V_{OGE} (\vect{r}_1 - \vect{r}_2) |\psi_0^{(1)}(\vect{r}_1^{\perp})
   |^2 |\psi_0^{(2)}(\vect{r}_2^{\perp})|^2 |\psi_0^{(z)}(\eta_z)|^2.
\end{equation}
In the momentum space one obtains
\begin{equation}
  \label{eq:28}
  \langle V_{OGE} \rangle = \frac{1}{(2\pi)^3} \int d^3 \veq V(\veq) F[|\psi_0^{(1)}|^2]_{-\veq_{\perp}} F[|\psi_0^{(2)}|^2]_{\veq_{\perp}} F[|\psi_0^{(z)}|^2]_{q_{z}}
\end{equation}
where $F[...]$ are Fourier images
\begin{equation}
  \label{eq:29}
  F[|\psi_0^{(i)}|^2]_{\vep^i_{\perp}} = \int d^2 \vex^i_{\perp} |\psi_0^{(i)}(\vex^i_{\perp})|^2e^{i(\vep^i_{\perp} \cdot \vex^i_{\perp})} =
 e^{-\frac{(\vep^i_{\perp} )^2}{4\omega_i^{(0)} \Omega_i}};
\end{equation}
\begin{equation}
  \label{eq:30}
  F[|\psi_0^{(z)}|^2]_{\pi_{z}} = \int d \eta_z |\psi_0^{(z)}(\eta_z)|^2e^{i p_z \eta_z} = e^{-\frac{\pi_z^2}{4 \tilde{\omega}^{(0)} \Omega_z}},
\end{equation}
where $\Omega_i$ and  $\Omega_z$ are given by (\ref{eq:17-2}). Comparing   this result in case of the neutral meson with the exact one, obtained with
Pseudomomentum procedure, one has to make a correction for the wave function, see Appendix B for details.

\section{Spin-dependent corrections}

A detailed review of the spin-dependent forces in PIH framework is given in
\cite{32}. Here we only emphasize that the spin-dependent perturbative
corrections arise from the $\langle \sigma_i F \cdot \sigma_j F \rangle$
correlators, where $\sigma_i$ are Clifford $4 \times 4$ $\sigma_{\mu \nu}$ for
i-th quark constituent and $F$ are non-abelian field strength tensors.

 Averaging over the stochastic gluonic background field, one has two
types of corrections   --  the self-energy  term for $i = j$ and color-magnetic
spin-spin interaction terms for $i \neq j$, where $i,\ j$ are quark numbers
\begin{equation}
  \label{eq:31}
  \begin{split}
  \Delta M_{SE} & = -\frac{4 \sigma}{3 \pi \omega_i^{(0)}};\\
 V_{SS}^{i j} & = \frac{8\pi \alpha_s^{(0)}}{9 \omega_i^{(0)} \omega_j^{(0)}} \delta(\ver_i - \ver_j) (\gvect{\sigma}_i \cdot \gvect{\sigma}_j).
  \end{split}
\end{equation}
The  self-energy correction $\Delta M_{SE}$  in (\ref{eq:31}) was used in a large
number of calculations \cite{33}, confirmed by the experimental data and
lattice simulations. In case of an external MF we retain  in $\Delta M_{SE}$
the value  $ \omega_i^{(0)} =\omega_i^{(0)}(eB = 0)$, instead of
$\omega_i^{(0)}(eB)$, which  does not change appreciably
 $ M_{\rm total}$.

A different story is for the spin-spin interaction in (\ref{eq:31}). As it was
shown in \cite{211,32,34}, the  wave function of hadronic and atomic systems
becomes ``focused" at the origin by MF, i.e. $|\Psi_0(0)|^2 \sim eB$ for large
MF value. This ``magnetic focusing" phenomenon could   induce  the
fall-to-the-center phenomenon for the lowest lying ZHS states. However,   as
shown in \cite{32}, the colormagnetic fields cannot violate the positivity of
the $q\bar q$ spectra, implying that some sort of the cut-off parameter must
 occur in the whole perturbative  series with nonperturbative background. Moreover,
 PIH method has a natural dimensional cutoff parameter for color field $\lambda
\simeq 1 \ GeV^{-1}$ -- correlation length of the vacuum gluonic background,
which    should  be used to  smear $\delta$-function in (\ref{eq:31}) \cite{32a}
\begin{equation}
  \label{eq:33}
  \delta(\ver) \rightarrow \frac{1}{\pi^{3/2}\lambda^3}e^{-\frac{\vect{r}^2}{\lambda^2}},
\end{equation}
and after the averaging with the CS meson wave function (\ref{eq:17}) one
obtains the  spin-spin matrix element
\begin{equation}
  \label{eq:34}
   \begin{split}
  \langle a_{SS} \rangle (\gvect{\sigma}_1 \cdot \gvect{\sigma}_2) = & \int V_{SS}^{12}|\Psi_0|^2 d^3 \ver_1 d^3 \ver_2 =\\
&  \frac{1}{\sqrt{\pi^3 \lambda^6}} \frac{\sqrt{\tilde{\omega}^{(0)} \Omega_z}}{\sqrt{\frac{1}{\lambda^2} + \tilde{\omega}^{(0)} \Omega_z}} \frac{1}{ 1 + \frac{1}{\lambda^2}\frac{\omega_1^{(0)} \Omega_1 + \omega_2^{(0)} \Omega_2}{\omega_1^{(0)} \Omega_1 \omega_2^{(0)} \Omega_2}} \frac{8\pi \alpha_s^{(0)}}{9 \omega_1^{(0)} \omega_2^{(0)}}  (\gvect{\sigma}_1 \cdot \gvect{\sigma}_2).
  \end{split}
\end{equation}
Smearing procedure prevents the collapse of   the meson in strong MF and it
stops the  unbounded fall of the total mass value in increasing MF.

It is important to notice here that the approximation of the confinement
potential by the harmonic oscillator potential (\ref{eq:6}) gives too small
value for $|\Psi_0(0)|^2$ and the hyperfine splitting $\Delta E = 4 \langle
a_{SS} \rangle$ between the non-chiral $\pi^-$ and $\rho^-$ mesons at $eB = 0$ is too
small (see Fig. 3 and 7) as compared with realisitc case of linear interaction. 
Moreover, the pion mass at $eB=0$ is additionally shifted down by chiral dynamics,
which we shall take into account in Section 7.

\section{Spin-isospin splittings in MF}

As pointed out in Section 1, MF violates spin and isospin  symmetries, therefore
$\pi^0,\ \rho^0$ split into 8 states and each $\pi^+,\ \rho^+$ and $\pi^-,\
\rho^-$ states split into 4 states in MF correspondingly. Using the asymptotics
(\ref{eq:24}), obtained in Section 3 for strong MF regime, one has
\begin{equation}
\label{eq:35}
  \begin{split}
1) \  \rho^+&(s_z = 1)  =  | u \uparrow \ \bar{d} \uparrow \rangle \ ZHS\\
2) \  \rho^+&(s_z = -1) =  | u \downarrow\ \bar{d} \downarrow \rangle \ II)\\
3) \  \rho^+&(s_z = 0)  =  \frac{1}{\sqrt{2}}
 \left(| u \uparrow \ \bar{d} \downarrow \rangle   + | u \downarrow \ \bar{d} \uparrow \rangle \right) \ I)\\
4) \  \pi^+&(s_z = 0)  =  \frac{1}{\sqrt{2}} \left(| u \uparrow \ \bar{d}
 \downarrow \rangle   - | u \downarrow \ \bar{d} \uparrow \rangle \right) \ I)\\
5) \  \rho^0&(s_z = 1)  =  \frac{1}{\sqrt{2}} \left(| u \uparrow \
\bar{u} \uparrow \rangle   + | d \uparrow \ \bar{d} \uparrow \rangle \right) \ I)\\
6) \  \rho^0&(s_z = -1)  =  \frac{1}{\sqrt{2}} \left(| u \downarrow \ \bar{u}
 \downarrow \rangle   + | d \downarrow \ \bar{d} \downarrow \rangle \right) \ I)\\
7) \  \rho^0&(s_z =  0)  =  \frac{1}{\sqrt{2}} \left[ \frac{1}{\sqrt{2}}
 \left(| u \uparrow \ \bar{u} \downarrow \rangle   + | d \uparrow \ \bar{d} \downarrow \rangle \right) +  \frac{1}{\sqrt{2}} \left(| u \downarrow \ \bar{u} \uparrow \rangle
  + | d \downarrow \ \bar{d} \uparrow \rangle \right) \right] \ ZHS + II)\\
8) \  \pi^0&(s_z =  0)   =  \frac{1}{\sqrt{2}} \left[ \frac{1}{\sqrt{2}} \left(| u
 \uparrow \ \bar{u} \downarrow \rangle   + | d \uparrow \ \bar{d} \downarrow \rangle \right)  - \frac{1}{\sqrt{2}} \left(| u \downarrow \ \bar{u} \uparrow \rangle   + | d \downarrow \
   \bar{d} \uparrow \rangle \right) \right] \ ZHS  + II)\\
9) \  \rho^-&(s_z = 1)  = | d \uparrow \ \bar{u} \uparrow \rangle \ II)\\
10) \  \rho^-&(s_z = -1) =  | d \downarrow\ \bar{u} \downarrow \rangle \ ZHS\\
11) \  \rho^-&(s_z = 0)  =  \frac{1}{\sqrt{2}} \left(| d \uparrow \ \bar{u}
 \downarrow \rangle   + | d \downarrow \ \bar{u} \uparrow \rangle \right) \ I)\\
12) \  \pi^-&(s_z = 0)  =  \frac{1}{\sqrt{2}} \left(| d \uparrow \ \bar{u}
\downarrow \rangle   - | d \downarrow \ \bar{u} \uparrow \rangle \right) \ I)
\end{split}
\end{equation}
Here on the l.h.s we have the standard spin-isospin configurations for mesons
at zero MF, and on the r.h.s we have asymptotic classification according to
({\ref{eq:24}) in strong MF for the corresponding states. 
The states 1)-4), 5)-8) and 9)-12) are composed of quarks and antiquarks in the combinations which yield the required 
spin and isospin values of $\pi$ and $\rho$ mesons at $eB=0$.

With increasing MF the eigenvalues of the total Hamiltonian (\ref{eq:5}) at nonzero MF demonstrate two types of phenomena: 
a) the mixing effect, due to spin-spin forces, equivalent to the Stern--Gerlach phenomenon, when the MF eigenstate can be expanded in two $eB=0$ eigenstates; b) the splitting effect, when the zero MF state composed of $u \bar u$ and $d \bar d$ components, splits into two trajectories due to isospin flavor. Finally, the trajectories for charged mesons like $\rho^+(s_z=1)$ and $\rho^+(s_z=-1)$   starting at the same mass at $eB=0$, split into two for $eB > 0$. 

To take into account the spin-spin interaction, we choose the basis states $|++ \rangle$, $|+- \rangle$, $|-+ \rangle$,  $|--
\rangle$ in spin space. The states 1) and 2), that corresponds to $\rho^+(s_z=1)$ and $\rho^+(s_z=-1)$ mesons at $eB=0$ correspondingly, are diagonal and their dynamical masses are 
\begin{equation}
  \begin{split}
  M_d^{++} &  = \langle ++ | H_d | ++ \rangle;\\
 M_d^{--} & =  \langle -- | H_d | -- \rangle.
  \end{split}
\end{equation}
After the stationary point analysis (\ref{eq:16}) one has two sets of parameters $(\omega_1^{++ \ (0)}, \omega_2^{++ \ (0)})$ and 
$(\omega_1^{-- \ (0)}, \omega_2^{-- \ (0)})$. The total mass of these states according to PIH formalism are given by
\begin{equation}
  \begin{split}
 &   M_{total} (\rho^+(s_z = 1)) = (M_d^{++} + \langle V_{OGE} \rangle + \Delta M_{SE} -  \langle a_{SS} \rangle )|_{\omega_1^{++ (0)},  \omega_2^{++ (0)}};\\
 &   M_{total} (\rho^+(s_z = -1)) = (M_d^{--} + \langle V_{OGE} \rangle + \Delta M_{SE} -  \langle a_{SS} \rangle )|_{\omega_1^{-- (0)},  \omega_2^{-- (0)}},
  \end{split}
\end{equation}
$M_{total}(\rho^+(s_z \pm 1))$ gives rise to two trajectories in MF starting at $\rho^+$ meson mass at $eB=0$. 

The behavior of states 3) and 4) corresponding to $\rho^+(s_z=0)$ and $\pi^+$ at zero MF is more complicated. These states are composed of $|u \downarrow \bar d \uparrow \rangle = |-+ \rangle$ and  $|u \uparrow \bar d \downarrow \rangle = |+- \rangle$ combinations at $eB=0$. When the MF increases, the states start to mix in the mutually orthogonal combinations 
\begin{equation}
  \pi^+, \rho^+(s_z=0) = \alpha \left( \frac{|u \uparrow \bar d \downarrow \rangle + |u \downarrow \bar d \uparrow \rangle}{\sqrt{2}} \right) + \beta \left( \frac{|u \uparrow \bar d \downarrow \rangle - |u \downarrow \bar d \uparrow \rangle}{\sqrt{2}} \right).
\end{equation}
The basis vectors are equal to the $\pi^+$ and $\rho^+(s_z=0)$ states at $eB=0$. The mixing phenomenon is defined by the non--diagonal spin--spin matrix elements
\begin{equation}
  \label{eq:38}
  \begin{split}
  a_{12} & = \langle +- | \langle a_{SS} \rangle (\gvect{\sigma_1} \cdot \gvect{\sigma_2})| -+ \rangle|_{\omega_1^{+- (0)},  \omega_2^{+- (0)}};\\
  a_{21} & = \langle -+ | \langle a_{SS} \rangle (\gvect{\sigma_1} \cdot \gvect{\sigma_2})| +- \rangle|_{\omega_1^{-+ (0)},  \omega_2^{-+ (0)}};\
\end{split}
\end{equation}
The dynamical masses and the parameters $(\omega_1^{+- (0)}, \omega_2^{+- (0)})$, $(\omega_1^{-+ (0)}, \omega_2^{-+ (0)})$ are defined by the stationary point analysis for the $M_d^{+-}, 
\ M_d^{-+}$ and the diagonal elements
\begin{equation}
  \begin{split}
     M_{total}^{11} & = (M_d^{+-} + \langle V_{OGE} \rangle + \Delta M_{SE} -  \langle a_{SS} \rangle )|_{\omega_1^{+- (0)},  \omega_2^{+- (0)}};\\
    M_{total}^{22} & = (M_d^{-+} + \langle V_{OGE} \rangle + \Delta M_{SE} -  \langle a_{SS} \rangle )|_{\omega_1^{-+ (0)},  \omega_2^{-+ (0)}}.
  \end{split}
\end{equation}
The final step is to diagonalize the total mass matrix 
\begin{equation}
  \label{eq:39}
\begin{bmatrix}
   M_{total}^{11} & 2a_{12}  \\
   2a_{21} & M_{total}^{22}  \\
 \end{bmatrix}
\end{equation}
and to calculate mixing coefficients 
\begin{equation}
\label{splc} 
\begin{split}
& \alpha,\beta (\rho^+(s_z=0)) = \frac{1}{\sqrt{2}} \frac{1 \pm \frac{M_{total}^{11} - E_1}{a_{12}}}{\sqrt{1 + \left(\frac{M_{total}^{11} - E_1}{a_{12}} \right)^2}};\\
&  \alpha,\beta (\pi^+) =  \frac{1}{\sqrt{2}} \frac{1 \pm \frac{M_{total}^{11} - E_2}{a_{12}}}{\sqrt{1 + \left(\frac{M_{total}^{11} - E_2}{a_{12}} \right)^2}}
\end{split}
\end{equation}
The eigenvalues for (\ref{eq:39}) are
\begin{equation}
\label{spl}
     E_{1,2}  = \frac{1}{2}(M_{total}^{11} + M_{total}^{22}) \pm \sqrt{ \left(\frac{M_{total}^{22} - M_{total}^{11}}{2} \right)^2 + 4 a_{12} a_{21} }.
\end{equation}
The trajectory $E_1$ with "+" sign in (\ref{spl}) starts from the $\rho^+(s_z=0)$ mass at $eB=0$ and grows with MF, and the trajectory $E_2$ with 
with "-" sign corresponds to the $\pi^+$ at zero MF. The states $E_1$ and $E_2$ are mixtures of $\pi^+$ and $\rho^+(s_z=0)$ at $eB \neq 0$ with mixing coefficients defined by (\ref{splc}). 

One can define the states 1)-4) with the same isospin structure as quartet $(\pi^+, \rho^+)$. States 9)-12) also form the quartet $(\pi^-, \rho^-)$ with the  dynamics in MF the same as for $(\pi^+, \rho^+)$ if one changes spins and charge signs to the opposite. The states 5)-8) are composed of $u \bar u$ and $d \bar d$ configurations in isospin. Since the relativistic Hamiltonian is diagonal in isospin, one can split these states into two independent quartets $(\pi^0, \rho^0)(u \bar u)$ and $(\pi^0, \rho^0)( \bar d)$. The diagonal state 5) splits into two trajectories $\rho^0(s_z = 1) = |u \uparrow \bar u \uparrow \rangle$ from quartet $(\pi^0, \rho^0)(u \bar u)$ and $\rho^0(s_z = 1) = |d \uparrow \bar d \uparrow \rangle$ from quartet $(\pi^0, \rho^0)(d \bar d)$ starting from the $\rho^0(s_z = 1)$ mass at zero MF. The same situation holds for the state 6). The states 7)-8) demonstrate the most complicated behaviour in MF: a)the double splitting in isospin to $u \bar u$ and $d \bar d$ trajectories; b) the mixing in spin $\alpha |u \uparrow \bar u \downarrow \rangle + \beta | u \downarrow \bar u \uparrow \rangle$ due to spin-spin matrix elements.

\section{Pion chiral degrees of freedom in MF}

Unlike $\rho$ mesons, the pions obey the chiral dynamics and therefore one
should take  into account how it changes under the influence of MF. This topic
was studied in \cite{24} and here we exploit  the results of \cite{24} for
neutral and charged pions. The most important feature of these results is that
the GMOR relations \cite{37} are kept valid for neutral  pions in arbitrary
strong MF, while they are violated for charged pions. At the same time at zero
and small  MF $(eB<f^2_\pi)$ the pion mass is  defined by GMOR relations both
in  the neutral and charged case, $m^2_\pi = \frac{m_q|\lan \bar q q \ran
|}{f^2_\pi}$.

This last dependence $m^2_\pi \sim O(m_q)$  defines the main difference between
chiral and nonchiral pion trajectories.

We start with the neutral pion case, with the standard GMOR relations.

\begin{equation}
  \label{eq:41}
  m_{\pi}^2 f_{\pi}^2 =  {\bar{m} }{ } | \langle u \bar{u} + d \bar{d} \rangle |; \ \bar{m} = \frac{m_u + m_d}{2},
\end{equation}
where $f^2_\pi$ is given in \cite{38,39},
\begin{equation}
  \label{eq:42}
  f_{\pi}^2 = N_c M^2(0)
   \sum_{n = 0}^{\infty} \left( \frac{\frac{1}{2}
    |\psi_{n,i}^{(+-)}(0)|^2}{(m_{n,i}^{(+-)})^3}
     + \frac{\frac{1}{2} |\psi_{n,i}^{(-+)}(0)|^2}{(m_{n,i}^{(-+)})^3}    \right),
\end{equation}
where $M(0)$  is the confining kernel, $M(0) = \sigma \lambda\cong 0.15$ GeV,  and $\langle u \bar{u} \rangle$, $\langle d \bar{d} \rangle$ are quark condensates in MF
\begin{equation}
  \label{eq:43}
  \langle q \bar{q} \rangle_i =  N_c M(0)
     \sum_{n= 0}^{\infty}\left( \frac{\frac{1}{2} |\psi_{n,i}^{(+-)}
     (0)|^2}{m_{n,i}^{(+-)}} + \frac{\frac{1}{2} |\psi_{n,i}^{(-+)}(0)|^2}{m_{n,i}^{(-+)}}
     \right).
\end{equation}
Here $(+-)$ and $(-+)$ are   individual quark's spin projections and e.g. $\psi_{n,i}^{(-+)}$ is the full 
set of $q \bar q$ non-chiral wave functions obtained with PIH formalism, $n$ is the radial quantum number, see \cite{38,39} for details.

Since $m^{-+}_{n,i}$ is  fast growing with $eB$, one can retain in the sums
(\ref{eq:42}),(\ref{eq:43}) only the $(+-)$ terms, and obtain as in \cite{24}
the asymptotic behavior of the $\pi^0$ mass as $m^2_{\pi^0} = \frac{\bar
m}{M(0)}(\bar m^{(+-)})^2$, where $\bar m^{(+-)}$ is  close to the lowest mass
$m_{n,i}$.

 It can be
seen that all formulae in derivation of the ECL in \cite{24} are diagonal in
isospin flavor, so one can write an  independent GMOR relation for each
$\pi^0(u \bar{u})$ and $\pi^0(d \bar{d})$ mesons, that should split in MF
according to (\ref{eq:35})
\begin{equation}
  \label{eq:44}
  \begin{split}
&  m_{\pi (q \bar{q})}^2 f_{\pi(q \bar{q})}^2 =  {m_q  }{ } | \langle q \bar{q} \rangle | ; \\
&  m_{\pi (q \bar{q})}^2  =  \frac{m_q  }{  M(0)} \left(m^{(+-)}_{(q\bar q)}\right)^2| \langle q \bar{q} \rangle
  |, ~~ q=u,d
  \end{split}
\end{equation}

The result of the calculation for $\pi^0(u \bar{u})$ (solid line) according to
(\ref{eq:44}) is shown in  the Fig.2.

\begin{figure}[t]
    \begin{center}
    \includegraphics[width=.7\textwidth]{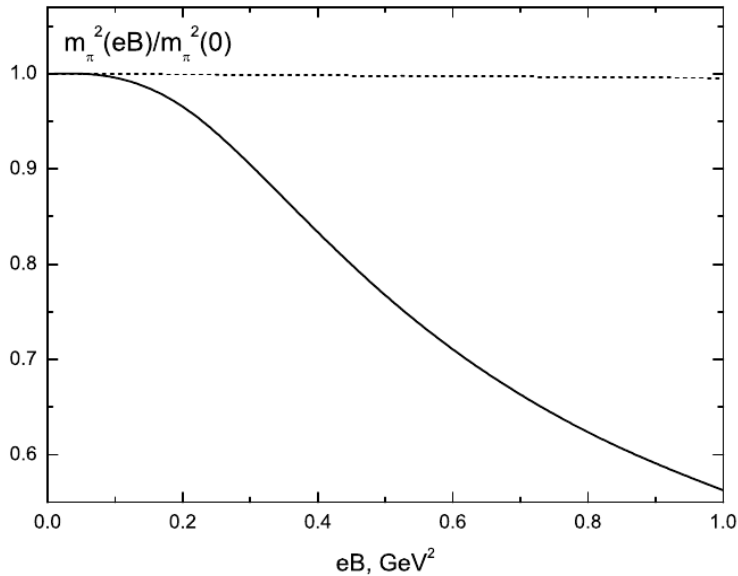}
    \caption{ Mass evolution of the chiral $\pi^0$ meson in MF from analytic PIH formalism.}
  \end{center}
  \end{figure}

\begin{figure}[t]
    \begin{center}
    \includegraphics[width=.8\textwidth]{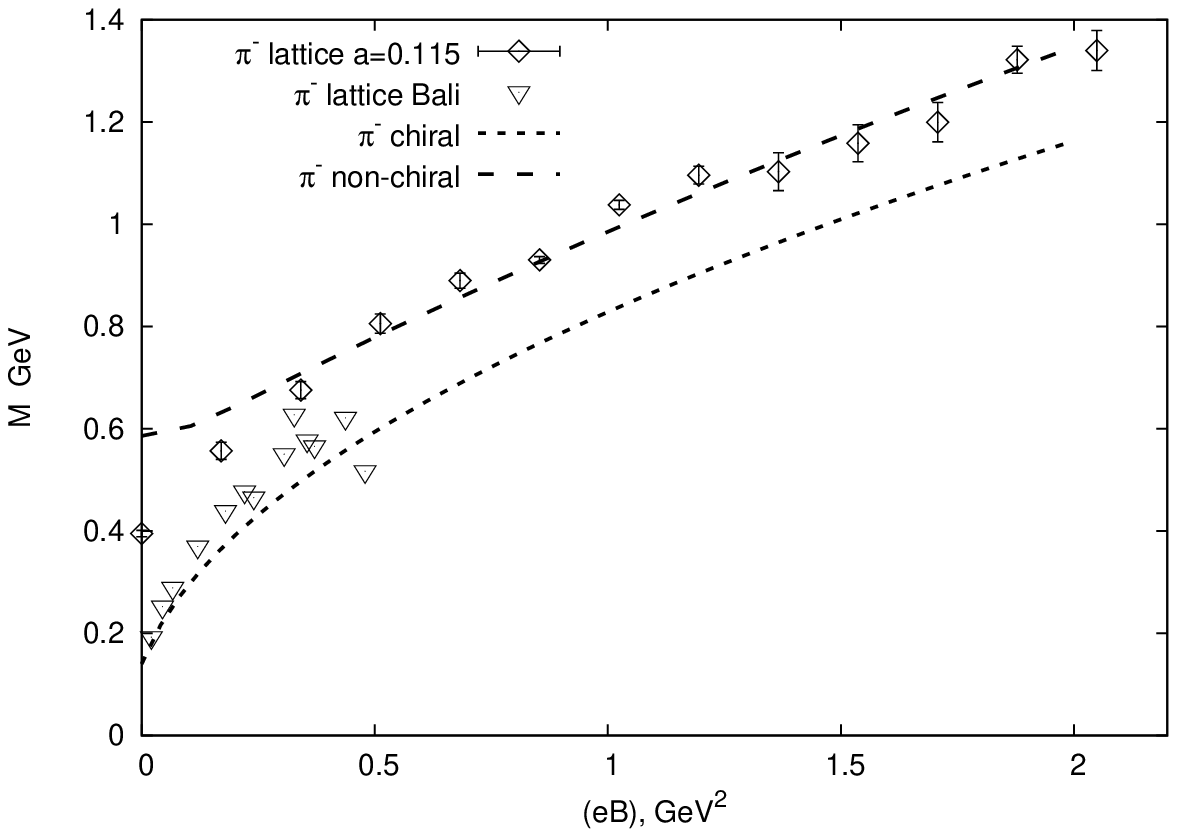}
    \caption{ Mass evolution of the chiral (solid line) and nonchiral (dashed line) $\pi^-$ meson in MF in comparison with the lattice data: triangles \cite{44} and quadrangles with lattice spacing $a=0.115 \ fm$ (present paper).}
  \end{center}
  \end{figure}

In the case of the charged pions, $\pi^+$ and $\pi^-$, the  situation is
drastically different, since they loose their chiral properties at large $eB>
\sigma$, and their asymptotics is  defined by the independent $u$ and $\bar d$
quarks (for $\pi^+)$, the  mode $I$ in (\ref{eq:24}), splitted in two trajectories, $M_{+-}(eB \gg \sigma)
\approx \sqrt{\frac23 eB}$ for $\pi^+$ and $M_{-+} (eB \gg \sigma) \approx \sqrt{\frac43 eB}$ for $\rho^+$ due to the $\pi^+-\rho^+$ mixing effect mentioned in Section 6. Taking into account the GMOR relations for $eB=0(eB < f^2_\pi)$, the asymptotics for the charged pions can be written as
\be M_{+-} (B) = \sqrt{m^2_\pi (0) +\frac23 eB},\label{44a}\ee 
where $m^2_\pi(0)$ is the pion mass at  $eB=0$.

In Fig.3 we plot the  trajectory of $M_{+-}(B)$ and our lattice
data together with the lattice data from \cite{44}.

One can summarize our method of chiral meson mass calculation as follows. First, one calculates the spectrum of non-chiral ("spectator") meson masses $m_{n,i}^{(+-)}$ in the MF. Second, one uses (\ref{eq:42}), (\ref{eq:43}) to obtain $f_{\pi}^2$ and $\langle q \bar{q} \rangle$ as functions of $eB$. Finally, one exploits GMOR relations to extract the resulting chiral mass dependence on $eB$. The formalism of (\ref{eq:42}), (\ref{eq:43}) was checked without MF in \cite{57-1,58-1}, the resulting MF dependence of $f_{\pi}^2$ and $\langle q \bar{q} \rangle$ was checked vs. lattice data in \cite{24, 59-1}.

Summarizing the results for neutral and charged pions one can say that our
theoretical predictions are supported by lattice data, and indeed charged and
neutral pions behave quite differently  at large MF, violating and not
violating respectively the GMOR relations.

\section{Lattice calculations}

The ground state energies of $\pi$ and $\rho$ mesons are calculated in SU(3) lattice gauge theory without dynamical quarks.
  Technical details  were presented in our previous work
  \cite{Luschevskaya:2015a,Luschevskaya:2015b}.
  We use 198-336 lattice gauge configurations on the lattice with spacing $a= 0.115 \ fm$ and 195 configurations  for  $a= 0.095 \ fm$  in volume $18^4$. Solving the   Dirac equation numerically
\begin{equation}
D \psi_k=i \lambda_k \psi_k, \  \ D=\gamma^{\mu} (\partial_{\mu}-iA_{\mu}),
\label{Dirac}
\end{equation}
we found eigenfunctions $\psi_k$ and eigenvalues $\lambda_k$ for a quark in the background gauge field $A_{\mu}$.  Two    types of quarks  $u$ and $d$ are considered, which are   degenerate  in mass. An abelian MF interacts with quarks, so $U(1)$ gauge field is introduced   into the lattice version of the Dirac operator $D$ ~\cite{Neuberger:1997}
 \begin{equation}
   \begin{split}
A_{\mu \, ij} & =  A^{SU(3)}_{\mu \, ij} + A_{\mu}^{B} \delta_{ij},\\
 A^B_{\mu}(x) & =\frac{B}{2} (x_1 \delta_{\mu,2}-x_2\delta_{\mu,1}).
\end{split}
\end{equation}
Quark fields obey periodic boundary conditions in space and antiperiodic boundary conditions in time. The MF is quantized in a finite lattice volume.  Its  value is determined by the following formula
\begin{equation}
eB=\frac{6\pi k}{(aL)^2},  \ k \in Z,
\label{quantization}
\end{equation}
where $e$ is the elementary charge.
Taking the average over the background field $A$ we introduce the correlators in coordinate space
\begin{equation}
\langle\psi^{\dagger}(x) O_1 \psi(x) \psi^{\dagger}(y) O_2 \psi(y)\rangle_A,
\label{lattice:correlator}
\end{equation}
where  $O_1, O_2=\gamma_5,\, \gamma_{\mu}$
are Dirac matrices, $\mu, \nu=1,..,4$ are Lorentz indices,
$x$ and $y$ are  lattice coordinates.

 We  performed the numerical Fourier transform of (\ref{lattice:correlator}) in spatial discrete coordinates
 and set $\langle\textbf{p}\rangle=0$ since we are interested in the meson ground state energy.
To obtain the masses we expand the correlation function $\tilde{C}(n_t)$  into the exponential series
\begin{equation}
 \langle \psi^{\dagger}(\textbf{0},n_t) O_1 \psi(\textbf{0},n_t) \psi^{\dagger}(\textbf{0},0) O_2 \psi(\textbf{0},0)\rangle_A = \sum_k\langle 0|O_1|k \rangle \langle k|O^{\dagger}_{2}|0 \rangle e^{-n_t a E_k}.
\label{sum}
 \end{equation}

When the lattice time $n_t$ is large, the main contribution to the correlator (\ref{sum}) comes from the ground state. Due to the periodic boundary conditions the correlator  has the following form
\begin{equation}
\tilde{C}_{fit}(n_t)=A_0 e^{-n_t a  E_0} + A_0 e^{-(N_T-n_t)  a E_0}= 2A_0 e^{-N_T a E_0/2} \cosh ((\frac{N_T}{2}-n_t) a E_0),
 \label{sum33}
\end{equation}
where  $A_0$ is a constant, $E_0$ is the ground state energy,  $a$ is the lattice spacing. We find the energy $E_0$, as a fit parameter,  fitting the lattice correlators by formula (\ref{sum33}).
In order to minimize the errors and to exclude the contribution of excited states we take various values of $n_t$ from the interval $5 \leq n_t \leq N_T-5$.
The energy of the charged pion is calculated from the correlation function
\begin{equation}
C^{\pi^{\pm}}=\langle \bar{\psi}_d(\vec{0}, n_t)\gamma_5 \psi_u(\vec{0}, n_t)  \bar{\psi}_u(\vec{0}, 0)\gamma_5 \psi_d(\vec{0}, 0) \rangle.
\label{pion+-}
\end{equation}

 Fig.3 shows the energy of a charged pion for the lattice volume  $18^4$, the lattice spacing   $0.115\ \fm$ and the quark masses  $17.13\ \Mev$, which corresponds to the pion mass   $m_{\pi}=395$ MeV at zero MF. The resultant energy increases with the MF value.
 Errors were obtained through the $\chi^2$ method.  
According to the exponential fall of the correlator \eqref{sum33} at large $E$, the absolute error of $E$ should grow with energy. However, we do not see this tendency clearly since different number of gauge configurations were utilized for different values of the MF. These numbers are shown in Table \ref{Table1} in case of the charged meson.  We observed an increase of the error for the  MF values $eB>1.2$ GeV$^2$, presumably due to the worst convergence of our numerical procedure at high MF values and small quark masses. In Fig.4 we show the correlation functions for different MF values for comparison. 
 
 \begin{table}[tbp]
 \begin{center}
\begin{tabular}{|c|r|r|r|}
\hline
$eB$, GeV$^2$  &  $N_{conf}$ & $E_{\pi^{\pm}}$, GeV    & Error     \\
\hline
$0$            & $245$  & $0.395 $   &   $0.006$         \\
\hline
 $0.171$       & $235$  & $0.557$    &   $0.017$           \\
\hline
 $0.341$       & $245$  & $0.675$    &   $0.017$           \\
\hline
$0.512$        & $311$  & $0.806$    &   $0.019$            \\
\hline
$0.683$        & $198$  & $0.890$    &   $0.015$           \\
\hline
$0.854$        & $241$  & $0.930$    &   $0.007$           \\
\hline
$1.024$       & $232$  & $1.038$    &   $0.009$           \\
\hline
$1.195$       & $238$  & $1.096$    &   $0.017$           \\
\hline
$1.366$       & $320$  & $1.103$    &   $0.037$           \\
\hline
$1.537$       & $333$  & $1.158$    &   $0.036$           \\
\hline
$1.707$       & $336$  & $1.200$    &   $0.039$           \\
\hline
$1.877$       & $248$  & $1.322$    &   $0.027$           \\
\hline
$2.049$       & $244$  & $1.340$    &   $0.039$           \\
\hline
$2.220$       & $249$  & $1.476$    &   $0.061$           \\
\hline
\end{tabular}
\end{center}
\caption{The values of  the $\pi^{\pm}$ energy, its errors and the number of lattice configurations, which were used for the calculations  at lattice volume $18^4$, lattice spacing $a=0.115$ fm, bare quark mass $17.13$ MeV and various MF values.}
\label{Table1}
 \end{table}
 
 \begin{figure}[t]
   \begin{center}
    \includegraphics[height=.8\textwidth,angle=-90]{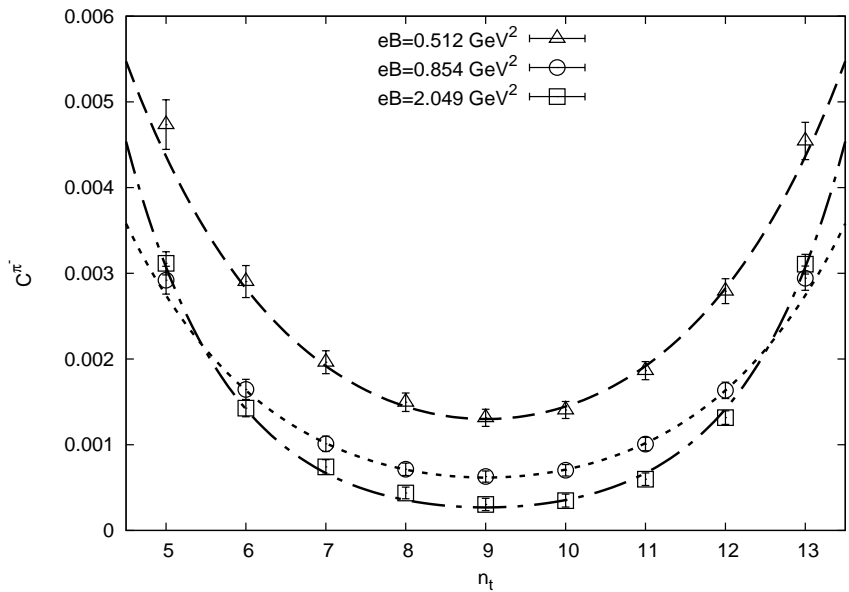}
    \caption{The correlation function \eqref{pion+-} versus $n_t$ for the lattice spacing $0.115$ fm, lattice
volume $18^4$ and bare quark masses $17.13$ MeV for several values of magnetic field $eB$.}
   \end{center}
\label{figCorr}
    \end{figure}

The energy of neutral pion was calculated using the correlation function
\begin{equation}
  \begin{split}
C^{\pi^0}= & (\langle \bar{\psi}_d(\vec{0}, n_t)\gamma_5 \psi_d(\vec{0}, n_t)  \bar{\psi}_d(\vec{0}, 0)\gamma_5 \psi_d(\vec{0}, 0) \rangle +\\
 &\langle \bar{\psi}_u(\vec{0}, n_t)\gamma_5 \psi_u(\vec{0}, n_t)  \bar{\psi}_u(\vec{0}, 0)\gamma_5 \psi_u(\vec{0}, 0) \rangle)/\sqrt2
  \end{split}
\label{pion0}
\end{equation}
 The resultant energies for the $\pi^0(u \bar u)$ and $\pi^0(d \bar d)$ configurations diminish with the increase of the MF as shown in Fig.5-6.

 The correlation functions of charged  $\rho$ mesons for three spatial directions are given by the following  relations
\begin{equation}
C_{xx}^{VV}=\langle \bar{\psi}_u(\textbf{0},n_t) \gamma_1 \psi_u(\textbf{0},n_t)
    \bar{\psi}_d(\textbf{0},0) \gamma_1 \psi_d(\textbf{0},0)\rangle,
    \label{Cxx}
\end{equation}
\begin{equation}
C_{yy}^{VV}=\langle \bar{\psi}_u(\textbf{0},n_t) \gamma_2 \psi_u(\textbf{0},n_t)
    \bar{\psi}_d(\textbf{0},0) \gamma_2 \psi_d(\textbf{0},0)\rangle,
        \label{Cyy}
\end{equation}
  \begin{equation}
 C_{zz}^{VV}=\langle \bar{\psi}_u(\textbf{0},n_t) \gamma_3 \psi_u(\textbf{0},n_t)
    \bar{\psi}_d(\textbf{0},0) \gamma_3 \psi_d(\textbf{0},0)\rangle.
    \label{Czz}
 \end{equation}
 If an abelian MF is directed along the 'z' axis, the $\rho$ meson energy with  $s_z=0$ spin projection  to the MF direction is determined by the $C_{zz}^{VV}$ correlator.
The ground state energies of the $\rho$ meson  with spin projections $s_z=+1$ and $s_z=-1$ are determined by the following combinations of correlators
    \begin{equation}
C^{VV}(s_z=\pm 1)= C^{VV}_{xx}+C^{VV}_{yy} \pm i(C^{VV}_{xy}-C^{VV}_{yx}).
\label{eq:CVV1}
    \end{equation}
 We have obtained that the energy of the $\rho^-$ meson with the spin projection $s_z=-1$ diminishes as a function of $eB$.
Fig.7 shows that the energies of $\rho^-$ meson with spin projections $s_z=0$ and $s_z=+1$ increase with the magnetic field value.
The energy of the neutral $\rho$ meson was calculated similarly to the charged $\rho$ meson, but in formulae \eq{Cxx}, \eq{Cyy} and \eqref{Czz} one has to consider the sum of the correlators for  $u$ and $d$ quarks.
In Fig.5-6 we represent the energy of neutral $\rho$ meson with various spin projections. The energies with $s_z=-1$
and $s_z=+1$ increase with the MF and coincide with each other.

\section{Results and discussion}
  \begin{figure}[t]
    \begin{center}
    \includegraphics[width=.8\textwidth]{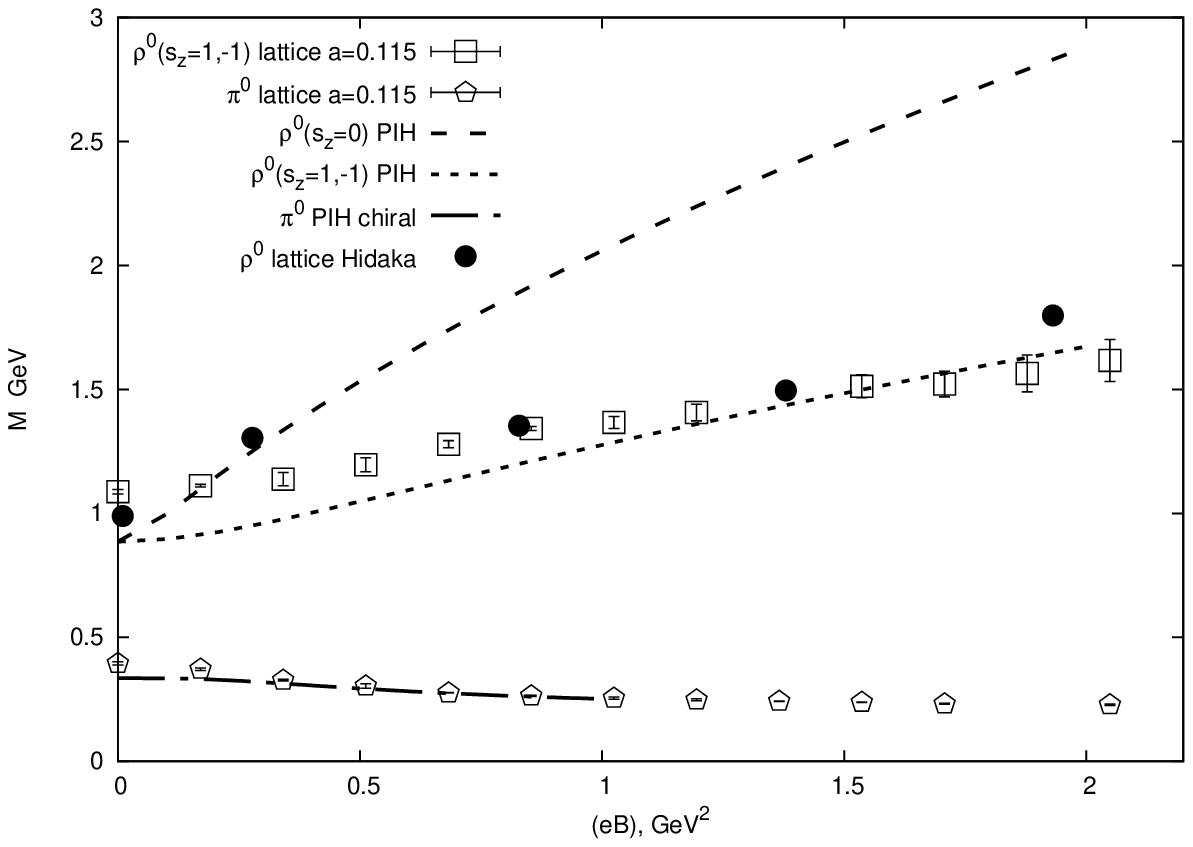}
    \caption{ Mass evolution of $(\pi^0, \ \rho^0)(u \bar u)$ quartet in MF from analytic (PIH) and lattice data (black circles are from \cite{43}).}
  \end{center}
  \end{figure}

  \begin{figure}[t]
    \begin{center}
    \includegraphics[width=.8\textwidth]{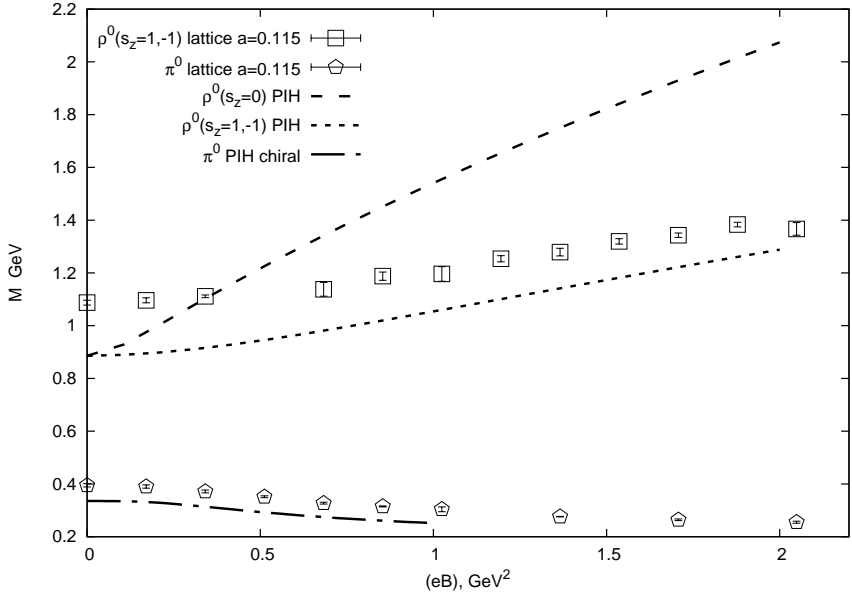}
    \caption{ Mass evolution of $(\pi^0, \ \rho^0)(d \bar d)$ quartet in MF from our analytic (PIH) and lattice data with $a=0.115 \ fm$.}
  \end{center}
  \end{figure}

  \begin{figure}[t]
    \begin{center}
    \includegraphics[width=.8\textwidth]{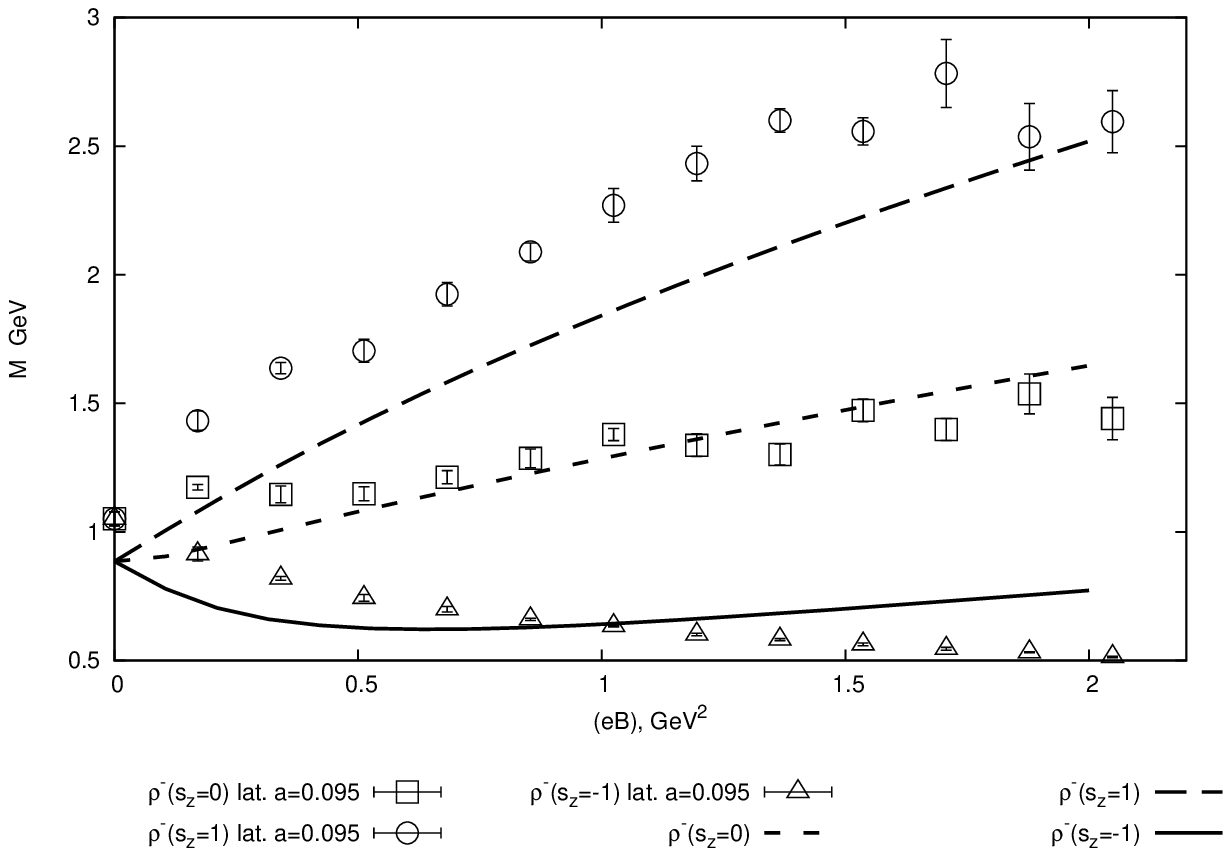}
    \caption{The $\rho^-$ meson mass evolution in MF from our analytic (PIH) and lattice data with $a = 0.095 \ fm$.}
  \end{center}
  \end{figure}

Our paper  contain analytic and numerical lattice results for all 12 $\pi,
\rho$  mass trajectories as functions of $eB$. The main  difficulty with
analytic results was  threefold. First of all, for charged states there is no
universal method of the separation of  c.m. and relative coordinates (unlike
the case of neutral mesons), and therefore we  have used a new special (however
approximate, $O(15\%)$) approach, called the CS  formalism. Secondly, strong MF
in lowest approximation brings in vacuum instability due to OGE forces and due
to  the hyperfine interaction, both growing fast with $eB$. We have eliminated
the OGE  instability  taking into account the screening effect \cite{21}, as
was shown in Section 4, see  also Fig.1. For the hyperfine problem we have used
the stability theorem of \cite{24}, ensuring the nonnegativity of hadron masses
in the  magnetic field, which implies that in higher order the combined nonperturbative
and perturbative effects must stabilize hyperfine interaction. To  this end we
have employed the smearing  radius of the hyperfine term of the  order of vacuum
correlation length $\lambda \simeq 0.2 \ fm $. Note, that this problem exists also
without MF and is usually solved in the same way.

Thirdly, the chiral dynamics, which governs pions at zero MF,  may be violated
by  MF,  and this was explicitly demonstrated in \cite{25}. Accordingly we  had
to consider separately  charged and neutral pions, where only the latter  keep
the  the chiral properties, see Fig.2 and 3.

Indeed we show in Fig.5 and 6 the behaviour of the $\pi^0(u \bar u)$ and $\pi^0(d \bar d)$ masses in MF, which follows
from the GMOR relation, where both $f_\pi(eB)$, and $\lan q \bar q \ran (eB)$ are
calculated via  non-chiral $q\bar q$ eigenvalues in MF. Note, that the chiral $\pi^0$ mass
in Fig.2 and the nonchiral $\pi^0$ mass are similar in behavior  but differ in scale. The latter is due to fact, that in chiral
dynamics $m^2_\pi$ is proportional to the quark mass $m_q$, see Eq.(\ref{eq:44}).

Moreover, the nonchiral neutral pion mass becomes negative for $eB > 0.6$ GeV,
when the standard hyperfine cut-off of $\lambda \approx 1 $ GeV$^{-1}$ is used,
which might require a smaller $\lambda$. This fact calls for an additional investigation.

 As for charged pions $\pi$,
one can see in Fig.3 a drastically different behavior which has growing
asymptotics of the type I according to (\ref{eq:24}) for chiral ($m^2_\pi=O(m_q^2)  $, lower curve) and
nonchiral $(m^2_\pi=O(\sigma)$ higher curve), cases. One can see in Fig.3 a
reasonable agreement of  lower curve with the lattice data of \cite{10} while
our  present lattice data in Fig.3 correspond to much larger $m_q$ and
therefore are shifted upwards.

Turning to the $\rho$ mass trajectories, one must remember our classification
in  Section 6, which implies, that both $\pi^0, \rho^0$ lines split into
$(u\bar u)$ and $(d\bar d)$ species and the growing trajectories are
proportional to $\sqrt{|e_q|B}$, yielding for those    a ratio equal to
$\sqrt{2}$.

The $\rho^0(s_z=1,-1)$ PIH trajectories in Fig.5 and 6 agree well with our
lattice data and with lattice data from \cite{43}, as well as $\pi^0$ ($u\bar u)$ and $\pi^0(d\bar d)$ trajectories.

Note the difficulty in lattice evaluation of the $\rho^0(s_z=0)$ lines which
mix with  the much lower $\pi^0$ trajectories.

A very interesting situation occurs for $\rho^-(\rho^+)$ mass trajectories,
presented in Fig.7. Only one of those belong to the ZHS type and tends to a
constant at large MF, and both lattice and analytic curve agree within our
accuracy $O(15\%)$, approximately the same kind of agreement is seen in Fig.7
for the trajectory of the type $I, \rho^- (s_z=0)$ and that of the type $II,
\rho^-(s_z=1)$. Summarizing, one can conclude, that our lattice data agree with
analytic predictions within our accuracy limits and our classification and theory based on the PIH
formalism for all $s$-wave $\pi, \rho$ mesons give a realistic physical picture
in this section.



\section*{Acknowledgements}
 The authors are grateful to O.V. Teryaev for discussion. 
The analytic calculations were provided by M.A.Andreichikov, B.O.Kerbikov and Yu.A.Simonov with support by the Russian Science Foundation grant 16-12-10414. The lattice calculations (Section VIII) are presented in this paper were provided by E.V.Luschevskaya and O.E.Solovjeva with support by the RScF grant 16-12-10059.

\appendix 

 \section{String tension renormalization in CS method}
The string tension renormalization procedure could be illustrated by an analogy from classical mechanics, where two point masses are connected by the spring with the following classical Lagrangian
\begin{equation}
  \label{eq:A1}
  L = \frac{m_1 \dot{x}_1^2}{2} + \frac{m_2 \dot{x}_2^2}{2} - \frac{k (x_1 - x_2)^2}{2}.
\end{equation}
In what follows we take substitution $m_1 = \omega_1,\ m_2 = \omega_2$ and $k = \frac{\sigma}{2\gamma}$, which makes the Lagrangian (\ref{eq:A1}) canonically conjugated to the relativistic Hamiltonian (\ref{eq:1}) with the confinement potential was taken in oscillator form (\ref{eq:6}) at $B=0$ up to momenta- and coordinate-independent terms. The Lagrangian could be canonically quantized in c.m. reference frame
\begin{equation}
  \label{eq:A2}
  E = \frac{P^2}{2M} + \frac{1}{2} \sqrt{\frac{k}{\mu}}(2n + 1);\ E_0 = \frac{1}{2} \sqrt{\frac{k}{\mu}},\ P = 0,
\end{equation}
where $E_0$ is ground state. On the other hand, one can describe the same system as two independent oscillators with opposite phases (for $P=0$), each of them is connected to the c.m. with its own spring, with Lagrangian
\begin{equation}
  \label{eq:A3}
   L = \frac{m_1 \dot{x}_1^2}{2} + \frac{m_2 \dot{x}_2^2}{2} - \frac{k_1 x_1^2}{2} - \frac{k_2 x_2^2}{2},
\end{equation}
where the stiffnesses are $\frac{k_1}{k_2} = \frac{m_1}{m_2}$. To proceed further with canonical quantization procedure, one has for ground the state energy for (\ref{eq:A3})
\begin{equation}
  \label{eq:A4}
  \begin{split}
  E =  \frac{P_z^2}{2M} +  & \frac{1}{2} \sqrt{\frac{k_1}{m_1}}(2n_1 + 1) + \frac{1}{2} \sqrt{\frac{k_2}{m_2}}(2n_2 + 1);\\
 E_0 = & \frac{1}{2} \sqrt{\frac{k_1}{m_1}} + \frac{1}{2} \sqrt{\frac{k_2}{m_2}};\ n_1 = n_2 = 0.
  \end{split}
\end{equation}
The expression (\ref{eq:A4}) should take into account that in the quantum case the phases of two harmonic oscillators should be entangled due to constraint $m_1x_1 +m_2x_2=0$. An explicit derivation of the "$\sigma$-renormalization" requires Dirac quantization formalism for constrained systems. Here we use an heuristic way, based on the correspondence principle. One can substitute the constraint $m_1x_1 +m_2x_2=0$ and $\frac{k_1}{k_2} = \frac{m_1}{m_2}$ to the Lagrangian (\ref{eq:A3}) with the result
\begin{equation}
 \label{eq:A5}
   L = L_1 = \frac{M}{m_2}\left(\frac{m_1\dot{x}_1^2}{2} - \frac{k_1 x_1^2}{2} \right) =  L_2 = \frac{M}{m_1}\left(\frac{m_2\dot{x}_2^2}{2} - \frac{k_2 x_2^2}{2} \right).
\end{equation}
In addition one could combine $L_1$ and $L_2$ to $L = \alpha L_1 + (1 - \alpha)L_2,\ \alpha \in [0,1]$ which has the same energy and preserves number of degrees of freedom. If one take $\alpha = \frac{1}{2}$, the conjugated to $L$ Hamiltonian is 
\begin{equation}
  \label{eq:A6}
  \begin{split}
  H = & p_1 \dot{x}_1 + p_2 \dot{x}_2 - L =\\ & \frac{2m_2}{M}\left(\frac{p_1^2}{2m_1} + \left(\frac{M}{2m_2} \right)^2 \frac{k_1 x_1^2}{2} \right) +  \frac{2m_1}{M} \left(\frac{p_2^2}{2m_2} + \left(\frac{M}{2m_1} \right)^2 \frac{k_2 x_2^2}{2} \right)
  \end{split}
\end{equation}
and after the canonical quantization procedure one has a ground state energy for (\ref{eq:A6})
\begin{equation}
  \label{eq:A7}
  E = \frac{2m_2}{M} \frac{1}{2} \sqrt{\left(\frac{M}{2m_2} \right)^2 \frac{k_1}{m_1}} + \frac{2m_1}{M} \frac{1}{2} \sqrt{\left(\frac{M}{2m_1} \right)^2 \frac{k_2}{m_2}}
\end{equation}
The ground state energy (\ref{eq:A7}) equals to (\ref{eq:A2}) if one redefines $k_1$ and $k_2$ as
\begin{equation}
  \label{eq:A8}
  k_1 = \frac{k}{1 + \frac{m_1}{m_2}};\ k_2 = \frac{k}{1 + \frac{m_2}{m_1}}.
\end{equation}
The resulting ground state energies, obtained with CS method and "$\sigma$-renormalization" procedure exactly coincide with the corresponding energies for the neutral mesons in the Pseudomomentum technique \cite{211} in a whole range of MF, see e.g. (\ref{eq:12}-\ref{eq:14}) in the main text of the paper.

\section{The wave function of the neutral meson in CS method}
One can suppose that the results for the averaged operator (\ref{eq:27}) for neutral mesons coincides with the exact one from \cite{211} with Pseudomomentum technique in strong MF regime ($eB \gg \sigma$) because of one-to-one correspondence for dynamical masses in Section 2. However, there is a difference about $30 \%$ (especially for ZHS states) because of the c.m. fixing procedure, i.e. the translational invariance breaking in CS method (see Section 2). The nature of this discrepancy is the lowest Landau level (LLL) degeneracy in angular momentum projection $m$ in symmetric gauge, when MF is strong enough to make the confinig force negligible. Let's consider the Hamiltonian for a single particle in MF to illustrate this statement
\begin{equation}
  \label{eq:B1}
  H = \frac{1}{2m}\left( \hat{\vect{p}} - \frac{e}{2} \vect{B} \times \vect{x} \right)^2.
\end{equation}
The corresponding spectrum for this Hamiltonian is
\begin{equation}
  \label{eq:B2}
  E = \Omega(2n + |m| - m +1),
\end{equation}
where $\Omega = \frac{eB}{2m}$ is cyclotron frequency, $n$ is oscillator quantum number and $m$ is an angular momentum projection to the direction of the MF. It's clear that there is an infinite degeneracy for LLL $n=0$ for $m = 0,1,2,...$. If we go to the complex coordinates $z = x+ iy$, the wave function for the ground state could be written as
\begin{equation}
  \label{eq:B3}
  \psi_0^{(m)} = \sum_{m}A_m(z^*)^m e^{-\frac{z^* z}{4}(eB)},
\end{equation}
where $A_m$ is an arbitrary constant. Extending this formalism to the case of two non-interacting particles with opposite charges in MF, one can write two-particle ground state wave function
\begin{equation}
  \label{eq:B4}
  \psi_0 = \sum_{m_1,m_2}A_{m_1,m_2}(z_1^*)^{m_1}(z_2)^{(m_2)} e^{-\frac{z_1^*z_1}{4}(eB) - \frac{z_2^*z_2}{4}(eB)},
\end{equation}
Here we also rewrite the exact wave function for the neutral meson from \cite{211} was got with Pseudomomentum procedure
\begin{equation}
  \label{eq:B5}
  \psi_0^{(p)} = \frac{1}{\sqrt{\pi^{3/2}r_{\perp}^2 r_z}} e^{-\frac{\eta_{\perp}^2}{2r_{\perp}^2} - \frac{\eta_z^2}{2r_z^2}},
\end{equation}
where radii and their asymptotics in $eB \rightarrow \infty$ regime are
\begin{equation}
  \label{eq:B6}
  \begin{split}
  r_{\perp} = & \sqrt{\frac{2}{eB}}\left(1 + \frac{4\sigma \tilde{\omega}^{(0)}}{\gamma^{(0)} (eB)^2} \right)^{-\frac{1}{4}} \rightarrow \sqrt{\frac{2}{eB}}, \\
 r_z = & \left(\frac{\gamma^{(0)}}{\sigma \tilde{\omega}^{(0)}} \right)^{\frac{1}{4}} \rightarrow \frac{1}{\sqrt{\sigma}}
  \end{split}
\end{equation}
Comparing transversal parts of the wave functions for CS method (\ref{eq:17-2}) as (f) and Pseudomomentum method as (p) in strong MF for ZHS meson
\begin{equation}
  \label{eq:B7}
  \begin{split}
& \psi_f^{\perp}(eB \gg \sigma) \sim e^{-\frac{eB}{4} \left[(\vect{x}_1^{\perp})^2 + (\vect{x}_2^{\perp})^2 \right]},\ vs\\
& \psi_p^{\perp}(eB \gg \sigma) \sim e^{-\frac{eB}{4}(\vect{x}_1^{\perp} - \vect{x}_2^{\perp})^2} = \psi_f(eB \gg \sigma) e^{-\frac{eB}{4}(\vect{x}_1 \cdot \vect{x}_2)}.
  \end{split}
\end{equation}
It's evident that the term $exp \left\{-\frac{eB}{4}(\vect{x}_1 \cdot \vect{x}_2) \right\} $  in (\ref{eq:B7}) is formed by the power series in $m_1$ and $m_2$ entering before the exponent in (\ref{eq:B4}). So, the difference between the CS wave function $\psi_f$ and the exact wave function $\psi_p$ is given by the superposition of the degenerate LLL basis wave functions. This additional term in (\ref{eq:B7}) gives about 30\% of the total value of the CS Coulomb integral  (\ref{eq:27}) for neutral mesons. The nature of this underestimation is clear - an additional term in  (\ref{eq:B4}) recovers translational invariance of the c.m. for the $\psi_f$ wave function. Also one should note that this correction doesn't exist for the charged meson case because of lack of the c.m. tranlational invariance due to c.m. precession in MF.
The final step is to add an additional multiplier $exp \left\{-\frac{eB}{4}(\vect{x}_1 \cdot \vect{x}_2) \right\} $ to our CS wave function (\ref{eq:17-2}) by hand according to previous speculations. This modification gives us 10\% accuracy for the Coulomb correction integral (\ref{eq:27}) in comparison with one was obtained in \cite{211}.

\end{document}